\documentclass[shortAfour,sageh,times]{sagej}

\usepackage{moreverb,url}

\usepackage[colorlinks,bookmarksopen,bookmarksnumbered,citecolor=red,urlcolor=red]{hyperref}

\newcommand\BibTeX{{\rmfamily B\kern-.05em \textsc{i\kern-.025em b}\kern-.08em
T\kern-.1667em\lower.7ex\hbox{E}\kern-.125emX}}

\def\volumeyear{2016}

\begin{document}

\runninghead{Sarkar:  Superlinear Cities, Finite Limits}

\title{Superlinear cities, finite limits: Three characteristic scales of urban agglomeration}

\author{Somwrita Sarkar\\[4pt]
\normalsize The University of Sydney, Australia\\
\normalsize\texttt{somwrita.sarkar@sydney.edu.au}}





\begin{abstract}

Urban scaling research has established that many socioeconomic outputs increase superlinearly with city population, implying rising per-capita output. Although urban scaling models have incorporated
infrastructural and dissipative costs, they have
generally not identified the finite population
scales implied by independently scaling benefits
and diseconomies. We connect the classical
urban-economic trade-off between agglomeration
benefits and costs to a generalized scaling
framework in which these quantities follow
distinct power laws. When diseconomies scale
more rapidly than benefits 
the model yields three ordered characteristic
populations: the first (smallest) maximizes net benefit
per capita; the second maximizes total net
benefit; and the third (largest), at which aggregate benefits
equal diseconomies. Their absolute values depend
on the magnitudes of benefits and costs, but
their relative separation depends only on the
scaling exponents. We illustrate these relationships using published
scaling estimates from several countries, and
examine aggregate and extensive--intensive scaling
across 102 Australian Significant Urban Areas.
Aggregate employee income, rental payments and
daily journey to work commuting distances for journeys 
scale superlinearly, although the aggregate
rent--income exponent difference is statistically
inconclusive. Mean rent per reporting
renter household and estimated mean distance
for daily commutes increase more rapidly
with city population than mean employee income
per earner. These findings reveal intensive
differences obscured by aggregate scaling.
The framework shows how superlinear agglomeration
benefits can coexist with still faster-scaling
diseconomies and produce finite characteristic
urban scales without requiring benefits
themselves to become sublinear.

\end{abstract}


\maketitle

\section{Introduction}
Why cities grow, and whether there are limits to the advantages associated with increasing city size, are foundational questions in urban science and urban economics. Larger cities concentrate people, firms, knowledge, infrastructure and opportunities, generating productivity advantages through sharing, matching, learning and other agglomeration mechanisms. But increasing urban concentration and size also generate familiar diseconomies through congestion, commuting, land and housing scarcity, pollution, crime, crowding and pressure on urban systems. The central theoretical question is therefore not simply whether larger cities produce more, but how the benefits and burdens generated by increasing population change relative to one another as cities become larger.

The finite or optimal city size question has a long history in urban economics. Classical models balance agglomeration economies against commuting, congestion, land and other spatial costs, generating equilibrium or efficient city sizes from the interaction of benefits and costs \citep{henderson1974sizes,arnott1979optimal}. Subsequent work has shown that the relevant size may depend on urban function, production structure, amenities, spatial organization, institutions and the criterion by which an urban outcome is evaluated \citep{capello2004beyond,camagni2013optimal}.

Urban scaling theory begins from a different empirical regularity. Across systems of cities, socioeconomic quantities such as income, economic output, innovation and social interaction frequently increase superlinearly with population, while quantities associated with physical infrastructure commonly increase sublinearly \citep{bettencourt2007growth,bettencourt2013origins}. In the standard scaling interpretation, larger cities therefore combine increasing socioeconomic output per person with economies of scale in infrastructure provision.

This combination does not by itself generate a population scale beyond which further concentration becomes disadvantageous. In the Bettencourt-West like scaling models, sublinear exponents for infrastructure is an economy of scale, not a diseconomy: if the amount of infrastructure required per person falls with population, it reinforces rather than counteracts the apparent advantages of urban concentration. Existing scaling approaches have developed measures of net urban benefit by balancing social output against dissipative or infrastructural requirements \citep{bettencourt2013origins,sugar2021urban}, while related work has examined positive and negative urban externalities and environmentally constrained outcomes \citep{webster2024tiebout,gomezlievano2025agglomeration,reitemeyer2025urban}. 

The countervailing processes relevant to finite urban scale, however, need not be infrastructure itself. They may instead be quantities whose burden is generated or intensified by concentration: housing expenditure, commuting requirements, congestion, environmental exposure or other costs that increase with urban population. The question considered here is therefore more specific: \textit{\textbf{can finite characteristic urban scales arise while agglomeration benefits themselves continue to scale superlinearly?}}

\subsection{Paper Contribution}

We address this question by representing agglomeration benefits and urban diseconomies as independently measurable scaling processes. 


We then ask what follows when the elasticity of diseconomies exceeds that of agglomeration benefits even though both quantities remain superlinear. This formulation connects classical urban economic models to scaling models, and then finally to the question of finite city size. A result of note is that finite urban scale  does not require agglomeration benefits to plateau, become sublinear, or decline. Aggregate benefits may continue increasing more than proportionately with city size while the balance between benefits and burdens progressively deteriorates.

This simple asymmetry generates three distinct characteristic population scales. The first maximises average, or per-capita, net benefit. The second maximises total net benefit and occurs where the marginal increase in aggregate benefit equals the marginal increase in aggregate diseconomy. The third is the zero-net boundary at which the levels of aggregate benefit and diseconomy become equal. When diseconomies exceed benefits, these scales are successively ordered. 

The distinction is consequential. Beyond the first scale, per-capita (average) net benefit is already declining even though total net benefit continues to increase. Beyond the second scale, both average and total net benefit decline, while net benefit nevertheless remains positive until the third scale. A city may therefore continue to generate substantial aggregate benefit well beyond the population at which further concentration ceases to improve average outcomes. Further, these three thresholds provide a policy-relevant insight: urban planning and governance structures may explicitly state which set of benefits are normatively chosen or prioritised in planning for population growth or city growth. 

A further property of the model is that the absolute locations of these scales depend on the magnitudes of the benefit and diseconomy processes, but their relative separation does not. The ratios between the scales are independent of the coefficients of productivity and diseconomy and are determined entirely by the scaling exponents. The relative geometry of the three characteristic scales can therefore be inferred from population elasticities without requiring the two processes to be expressed in a common monetary unit. This provides a direct link between empirically estimated urban scaling relationships and the classical question of finite city size, while also showing why there is no unique ``optimal'' population without first specifying whether the criterion is average outcome, aggregate outcome or the boundary of positive net benefit.

We empirically apply the framework to test model implications on prior published results on scaling of benefits and diseconomies from several countries, providing estimates of the three scales in correspondence to current world population and largest city sizes. 

We also perform an extensive-intensive decomposition of scaling exponents on Australian data. Rather than treating aggregate relationships alone as evidence of diseconomies, we decompose each quantity into an extensive component (the number of people, households or commuters exposed to the process) and an intensive component measuring the outcome experienced by each relevant unit. Employee income is used as an observable indicator of the economic returns associated with urban scale. Housing burden is measured from Census rent distributions, while transport burdens are measured from Census journey-to-work distances, including separate analysis of private-motorized commuting. Across these domains, not only does the aggregate, but the intensive burden associated with urban scale rises more rapidly than intensive benefit. Mean rent increases systematically with city population, while mean commuting distance and mean private-motorised commuting distance increase still more strongly. The result therefore cannot be explained simply by the fact that larger cities contain more workers, renters, commuters or vehicles in aggregate: city size is associated with an increase in the burden experienced by the individual renter or commuter.

These empirical relationships are not interpreted as a complete social welfare accounting of urban benefits and costs. Rent, for example, is an expenditure burden to tenants but also income to property owners, while commuting distance captures only part of the generalized cost of travel. Their purpose is instead to test the key empirical premise of the theoretical model: whether economically relevant burdens can possess population elasticities exceeding those of observable agglomeration-related benefits, and whether this difference survives decomposition into extensive and intensive margins. The Australian results show that they can.

The framework is deliberately comparative-static. Cross-sectional scaling relationships describe systematic differences among cities of different population sizes and are not assumed to represent the temporal growth trajectory of an individual city \cite{MarquisBarthelemy2026}. The results establish a simple connection between two traditions that have largely posed the city-size problem differently. Urban economics has long shown that competing agglomeration benefits and urban costs can imply finite city sizes; urban scaling provides an empirical language in which the population elasticities of those processes can be estimated across systems of cities. Allowing benefits and diseconomies to scale independently shows that persistent superlinear agglomeration returns are entirely compatible with finite characteristic urban scales, and that the relationship among those scales can be derived directly from the scaling exponents themselves.

\section{Methods}
\label{sec:methods}
\subsection*{Urban diseconomies must scale faster than agglomeration benefits for a finite maximum in per-capita urban benefit}

Let aggregate urban benefits and diseconomies scale with population $N$ as
\begin{equation}
B(N)=AN^\beta
\end{equation}
and
\begin{equation}
D(N)=CN^\gamma,
\end{equation}
where $A,C>0$. The exponents $\beta$ and $\gamma$ describe the population elasticities of aggregate benefits and diseconomies, respectively. We consider the case in which both quantities scale superlinearly, such that $\beta>1$ and $\gamma>1$. Thus, neither increasing agglomeration returns nor increasing diseconomies is assumed to disappear as city size increases.

Average, or per-capita, net urban benefit is
\begin{equation}
\psi(N)
=
\frac{B(N)-D(N)}{N}
=
AN^{\beta-1}
-
CN^{\gamma-1}.
\label{eq:netbenefit}
\end{equation}

Since empirical evidence generally places both $\beta$ and $\gamma$ between 1 and 2, their corresponding per-capita exponents satisfy $0 < \beta-1 < 1$ and $0 < \gamma-1 < 1$. Thus, both per-capita urban benefits and diseconomies increase sublinearly with population. However, the net per-capita benefit, being the difference between these two terms, need not increase monotonically and may eventually
decline when $\gamma > \beta$.

To identify the population at which per-capita net urban benefit is maximized, we differentiate $\psi(N)$ with respect to population. The resulting derivative represents the marginal change in per-capita net urban benefit associated with an increase in
city population:
\begin{equation}
\psi'(N)
=
A(\beta-1)N^{\beta-2}
-
C(\gamma-1)N^{\gamma-2}.
\label{eq:firstderivative}
\end{equation}

Setting $\psi'(N)=0$ gives
\begin{equation}
A(\beta-1)N^{\beta-2}
=
C(\gamma-1)N^{\gamma-2},
\label{eq:firstderivative1}
\end{equation}
and therefore
\begin{equation}
N^{\gamma-\beta}
=
\frac{A(\beta-1)}
{C(\gamma-1)}.
\end{equation}

For $\gamma\neq\beta$, the stationary population is consequently
\begin{equation}
N_A
=
\left[
\frac{A(\beta-1)}
{C(\gamma-1)}
\right]^{\frac{1}{\gamma-\beta}}.
\label{eq:NA}
\end{equation}

We denote this population by $N_A$ because, when the stationary point is a maximum, it maximizes per-capita benefit rather than total net benefit.

Whether $N_A$ is a maximum or a minimum depends on the sign
of the second derivative. Differentiating again,
\begin{equation}
\psi''(N)
=
A(\beta-1)(\beta-2)N^{\beta-3}
-
C(\gamma-1)(\gamma-2)N^{\gamma-3}.
\end{equation}

Dividing both sides of Eq.~\ref{eq:firstderivative1} by $N$ yields
\begin{equation}
A(\beta-1)N^{\beta-3}
=
C(\gamma-1)N^{\gamma-3}.
\end{equation}

Substituting this equality into the expression for
the second derivative, evaluated at $N_A$, gives:
\begin{align}
\psi''(N_A)
&=
A(\beta-1)N_A^{\beta-3}
\left[(\beta-2)-(\gamma-2)\right] \nonumber\\
&=
A(\beta-1)N_A^{\beta-3}(\beta-\gamma).
\label{eq:curvature}
\end{align}

Because $A>0$, $\beta>1$, and $N_A$, all factors
in Eq.~\ref{eq:curvature} are strictly positive except
potentially $(\beta-\gamma)$. Consequently, the
second derivative is negative when $\gamma>\beta$,
confirming that $N_A$ is a local maximum rather
than a minimum under this condition. A finite maximum in per-capita
net urban benefit therefore requires
\begin{equation}
\gamma>\beta.
\label{eq:condition}
\end{equation}

Superlinear diseconomies are therefore not sufficient to generate a finite maximum in per-capita net benefit. Their population elasticity must exceed that of agglomeration benefits. When $\gamma>\beta$, both aggregate benefits and aggregate diseconomies continue to increase superlinearly with population. Their marginal quantities also continue to increase, but marginal diseconomies rise more rapidly than marginal benefits because
\begin{equation}
B'(N)=A\beta N^{\beta-1},
\qquad
D'(N)=C\gamma N^{\gamma-1},
\end{equation}
with $\gamma-1>\beta-1$. The marginal net return to additional population therefore progressively declines, and per-capita net benefit reaches a maximum at $N_A$ before subsequently falling. A finite characteristic scale can thus emerge even though agglomeration benefits remain superlinear and the marginal benefit of additional population continues to increase; neither saturation nor declining agglomeration benefits is required.

When $\gamma<\beta$, the stationary point is instead a minimum: benefits ultimately grow more rapidly than diseconomies and no finite maximum in per-capita net benefit exists. When $\gamma=\beta$, Eq.~\ref{eq:netbenefit} reduces to
\begin{equation}
\psi(N)=(A-C)N^{\beta-1},
\end{equation}
which is monotonic unless $A=C$, in which case net benefit is identically zero.

The condition $\gamma>\beta$ has a further consequence. The population $N_A$ is not the only finite scale generated by the model. Per-capita net benefit, total net benefit, and the equality of total benefits and diseconomies correspond to different mathematical conditions. The same two scaling relationships therefore generate three distinct characteristic urban scales, whose ordering and relative separation we derive next.

\subsection{Per-capita, total, and break-even urban scales are distinct, while their relative separation is independent of the coefficients A and C}

\textbf{The population that maximises per-capita net benefit precedes the population that maximises total net benefit, which in turn precedes the population at which benefits are fully offset by diseconomies.}

The first characteristic scale, derived above, is the population $N_A$ at which per-capita net benefit is maximised:
\begin{equation}
N_A
=
\left[
\frac{A(\beta-1)}
{C(\gamma-1)}
\right]^{\frac{1}{\gamma-\beta}}.
\label{eq:NA_repeat}
\end{equation}

A second characteristic scale follows from total net urban benefit,
\begin{equation}
T(N)
=
B(N)-D(N)
=
AN^\beta-CN^\gamma.
\label{eq:totalnet}
\end{equation}

Differentiating with respect to population gives
\begin{equation}
T'(N)
=
A\beta N^{\beta-1}
-
C\gamma N^{\gamma-1}.
\end{equation}

The stationary population therefore satisfies
\begin{equation}
A\beta N_T^{\beta-1}
=
C\gamma N_T^{\gamma-1},
\end{equation}
or
\begin{equation}
N_T
=
\left(
\frac{A\beta}
{C\gamma}
\right)^{\frac{1}{\gamma-\beta}}.
\label{eq:NT}
\end{equation}

Whether $N_T$ represents a maximum or a minimum follows from the second derivative of total net benefit:
\begin{equation}
T''(N)
=
A\beta(\beta-1)N^{\beta-2}
-
C\gamma(\gamma-1)N^{\gamma-2}.
\end{equation}

At $N_T$, the first-order condition implies
\begin{equation}
A\beta N_T^{\beta-2}
=
C\gamma N_T^{\gamma-2},
\end{equation}
after dividing both sides by $N_T$. Therefore,
\begin{align}
T''(N_T)
&=
A\beta N_T^{\beta-2}
\left[(\beta-1)-(\gamma-1)\right] \nonumber\\
&=
A\beta N_T^{\beta-2}(\beta-\gamma).
\label{eq:NTcurvature}
\end{align}

Since $A>0$, $\beta>1$, and $N_T>0$, the sign of $T''(N_T)$ is determined entirely by $(\beta-\gamma)$. Under the condition $\gamma>\beta$ established above,
\begin{equation}
T''(N_T)<0,
\end{equation}
so $N_T$ is a finite maximum of total net benefit.

At $N_T$, the marginal increase in aggregate benefit is exactly equal to the marginal increase in aggregate diseconomy. Under $\gamma>\beta$, this stationary point is a maximum of total net benefit. Thus $N_T$ differs conceptually from $N_A$: at $N_A$, per-capita net benefit has begun to decline, while total net benefit may still be increasing.

A third scale is the break-even population $N_0$, at which aggregate benefits and aggregate diseconomies are equal:
\begin{equation}
AN_0^\beta
=
CN_0^\gamma.
\end{equation}
Equivalently, per-capita benefit and diseconomy are equal at the same population. Solving gives
\begin{equation}
N_0
=
\left(
\frac{A}{C}
\right)^{\frac{1}{\gamma-\beta}}.
\label{eq:N0}
\end{equation}

The three characteristic populations have a strict ordering when $\gamma>\beta>1$. Comparing $N_A$ and $N_T$ gives
\begin{equation}
\frac{N_T}{N_A}
=
\left[
\frac{\beta(\gamma-1)}
{\gamma(\beta-1)}
\right]^{\frac{1}{\gamma-\beta}}.
\label{eq:NTratioNA}
\end{equation}

Because
\begin{equation}
\beta(\gamma-1)
-
\gamma(\beta-1)
=
\gamma-\beta
>
0,
\end{equation}
the quantity inside the brackets in Eq.~\ref{eq:NTratioNA} exceeds unity. Since $1/(\gamma-\beta)>0$,
\begin{equation}
N_T>N_A.
\end{equation}

Similarly,
\begin{equation}
\frac{N_0}{N_T}
=
\left(
\frac{\gamma}{\beta}
\right)^{\frac{1}{\gamma-\beta}}.
\label{eq:N0ratioNT}
\end{equation}

Since $\gamma>\beta$, this ratio is also greater than unity, and therefore
\begin{equation}
N_0>N_T.
\end{equation}

Hence,
\begin{equation}
N_A<N_T<N_0.
\label{eq:ordering}
\end{equation}

The ordering divides urban scale into distinct regimes. For $N<N_A$, both per-capita and total net benefit increase with population. Between $N_A$ and $N_T$, per-capita net benefit declines even though total net benefit continues to increase. Between $N_T$ and $N_0$, both per-capita and total net benefit decline, but net benefit remains positive. Only beyond $N_0$ do diseconomies exceed benefits and net benefit become negative.

Importantly, although the absolute locations of $N_A$, $N_T$, and $N_0$ depend on the coefficients $A$ and $C$, their relative separation does not. In addition to Eqs.~\ref{eq:NTratioNA} and \ref{eq:N0ratioNT},
\begin{equation}
\frac{N_0}{N_A}
=
\left[
\frac{\gamma-1}
{\beta-1}
\right]^{\frac{1}{\gamma-\beta}}.
\label{eq:N0ratioNA}
\end{equation}

The coefficients $A$ and $C$ cancel from all three expressions. The relative geometry of the three characteristic urban scales is therefore determined entirely by the scaling exponents $\beta$ and $\gamma$. This has an important implication: the model does not imply a universal absolute ``optimal'' city size. The absolute locations of $N_A$, $N_T$, and $N_0$ depend on the relative magnitudes of benefits and diseconomies through $A/C$ and may therefore vary substantially across urban systems. What is invariant is their relative ordering and separation. For a given pair of scaling exponents, systems with very different absolute population scales nevertheless share the same proportional relationship among the per-capita optimum, the total-net-benefit optimum, and the break-even scale.

\subsection{The three characteristic urban scales shift proportionally with the relative magnitude of benefits and diseconomies}

The absolute locations of the three characteristic urban scales depend systematically on the relative magnitude of agglomeration benefits and urban diseconomies, while their relative separation remains unchanged.

Recall that
\begin{align}
N_A
&=
\left[
\frac{A(\beta-1)}
{C(\gamma-1)}
\right]^{\frac{1}{\gamma-\beta}},
\\
N_T
&=
\left[
\frac{A\beta}
{C\gamma}
\right]^{\frac{1}{\gamma-\beta}},
\\
N_0
&=
\left(
\frac{A}{C}
\right)^{\frac{1}{\gamma-\beta}},
\end{align}
with $\gamma>\beta$.

Taking logarithms shows that each scale contains the common term
\begin{equation}
\frac{\ln A-\ln C}{\gamma-\beta}.
\end{equation}
Consequently, for each $N_j\in\{N_A,N_T,N_0\}$,
\begin{equation}
\frac{\partial\ln N_j}{\partial\ln A}
=
\frac{1}{\gamma-\beta}
>0,
\end{equation}
while
\begin{equation}
\frac{\partial\ln N_j}{\partial\ln C}
=
-\frac{1}{\gamma-\beta}
<0.
\end{equation}

Holding the scaling exponents constant, a proportional increase in the magnitude of agglomeration benefits therefore shifts all three characteristic scales towards larger populations by the same proportion. A proportional increase in the magnitude of diseconomies shifts all three towards smaller populations by the same proportion. These responses become stronger as the difference between the scaling exponents, $\gamma-\beta$, becomes smaller.

Because the elasticities are identical for $N_A$, $N_T$, and $N_0$, changes in $A$ or $C$ alter their absolute locations but not the ratios among them. The entire three-scale configuration therefore shifts along the population axis without changing its relative geometry. Cities or urban systems characterised by identical scaling exponents may consequently have very different absolute characteristic population scales if the magnitudes of their benefits and diseconomies differ, while preserving the same proportional separation between $N_A$, $N_T$, and $N_0$.

\subsection*{The per-capita and total net-benefit maxima are unique and occur before break-even}

When $\gamma>\beta>1$, $N_A$ and $N_T$ are the unique global maxima of per-capita and total net benefit, respectively, and both occur while net benefit remains positive.

For per-capita net benefit,
\begin{equation}
\psi(N)
=
AN^{\beta-1}
-
CN^{\gamma-1},
\end{equation}
the first derivative can be written as
\begin{equation}
\psi'(N)
=
N^{\beta-2}
\left[
A(\beta-1)
-
C(\gamma-1)N^{\gamma-\beta}
\right].
\label{eq:psi_derivative_factor}
\end{equation}

For $N>0$, the factor $N^{\beta-2}$ is positive. When $\gamma>\beta$, the expression in brackets decreases monotonically with population: it approaches $A(\beta-1)>0$ as $N\rightarrow0^+$ and tends to $-\infty$ as $N\rightarrow\infty$. It therefore crosses zero exactly once, at $N_A$. Hence
\begin{equation}
\psi'(N)>0 \quad \text{for } N<N_A,
\end{equation}
and
\begin{equation}
\psi'(N)<0 \quad \text{for } N>N_A.
\end{equation}
Thus $N_A$ is the unique global maximum of per-capita net benefit.

Per-capita net benefit is also strictly positive at this population. The first-order condition at $N_A$ implies
\begin{equation}
C(\gamma-1)N_A^{\gamma-1}
=
A(\beta-1)N_A^{\beta-1},
\end{equation}
and therefore
\begin{align}
\psi(N_A)
&=
AN_A^{\beta-1}
-
CN_A^{\gamma-1}
\nonumber\\
&=
AN_A^{\beta-1}
\left[
1-
\frac{\beta-1}{\gamma-1}
\right]
\nonumber\\
&=
AN_A^{\beta-1}
\frac{\gamma-\beta}{\gamma-1}
>0.
\label{eq:psi_NA_positive}
\end{align}

The same argument applies to total net benefit,
\begin{equation}
T(N)=AN^\beta-CN^\gamma.
\end{equation}
Its derivative can be written as
\begin{equation}
T'(N)
=
N^{\beta-1}
\left[
A\beta
-
C\gamma N^{\gamma-\beta}
\right].
\label{eq:T_derivative_factor}
\end{equation}

Again, the first factor is positive for $N>0$, while the expression in brackets decreases monotonically from $A\beta>0$ to $-\infty$. It crosses zero exactly once, at $N_T$. Hence total net benefit increases for $N<N_T$ and decreases for $N>N_T$, establishing $N_T$ as its unique global maximum.

Using the first-order condition at $N_T$,
\begin{equation}
C\gamma N_T^\gamma
=
A\beta N_T^\beta,
\end{equation}
gives
\begin{align}
T(N_T)
&=
AN_T^\beta
-
CN_T^\gamma
\nonumber\\
&=
AN_T^\beta
\left(
1-\frac{\beta}{\gamma}
\right)
\nonumber\\
&=
AN_T^\beta
\frac{\gamma-\beta}{\gamma}
>0.
\label{eq:T_NT_positive}
\end{align}

Thus neither maximum is a break-even point. Per-capita net benefit reaches its unique maximum at $N_A$ while total net benefit is still increasing, and total net benefit reaches its unique maximum at $N_T$ while benefits still exceed diseconomies. Only at the larger population $N_0$ does net benefit fall to zero. Together with the ordering established above,
$N_A<N_T<N_0$,
these results show that the decline in per-capita outcomes, the decline in total net benefit, and the complete exhaustion of net benefit occur at three distinct stages of urban scale.

\subsection*{The three characteristic scales define distinct regimes of urban benefit}

\begin{figure}[htbp]
    \centering
    \includegraphics[width=0.82\linewidth]{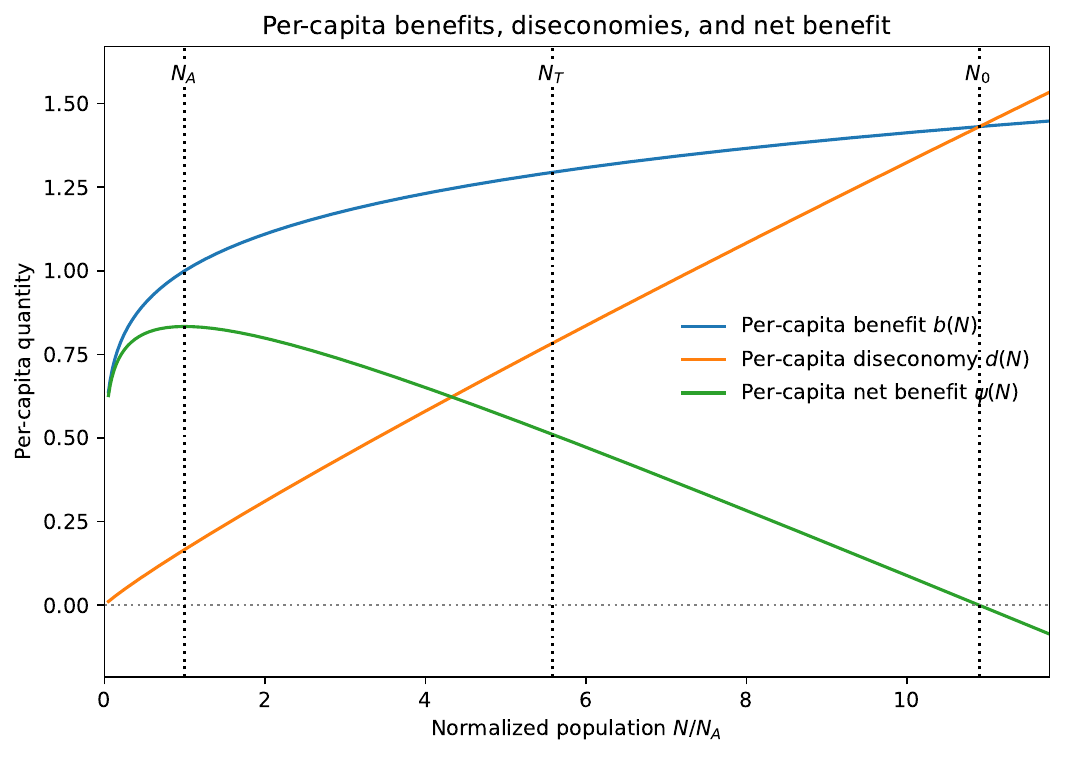}
    \caption{\textbf{Per-capita agglomeration benefits, diseconomies, and net benefit across the three characteristic urban scales.}
    Both per-capita benefits, $b(N)=AN^{\beta-1}$, and per-capita diseconomies, $d(N)=CN^{\gamma-1}$, increase with population. Because $\gamma>\beta$, diseconomies increase more rapidly. Per-capita net benefit, $\psi(N)=b(N)-d(N)$, reaches its unique maximum at $N_A$, declines thereafter, and reaches zero only at the larger break-even population $N_0$. The intermediate scale $N_T$ marks the population at which total, rather than per-capita, net benefit is maximised. Population is normalised by $N_A$. The illustration uses $\beta=1.15$ and $\gamma=1.90$, giving $N_T/N_A\simeq5.58$ and $N_0/N_A\simeq10.90$.}
    \label{fig:percapita_three}
\end{figure}

\begin{figure}[htbp]
    \centering
    \includegraphics[width=0.82\linewidth]{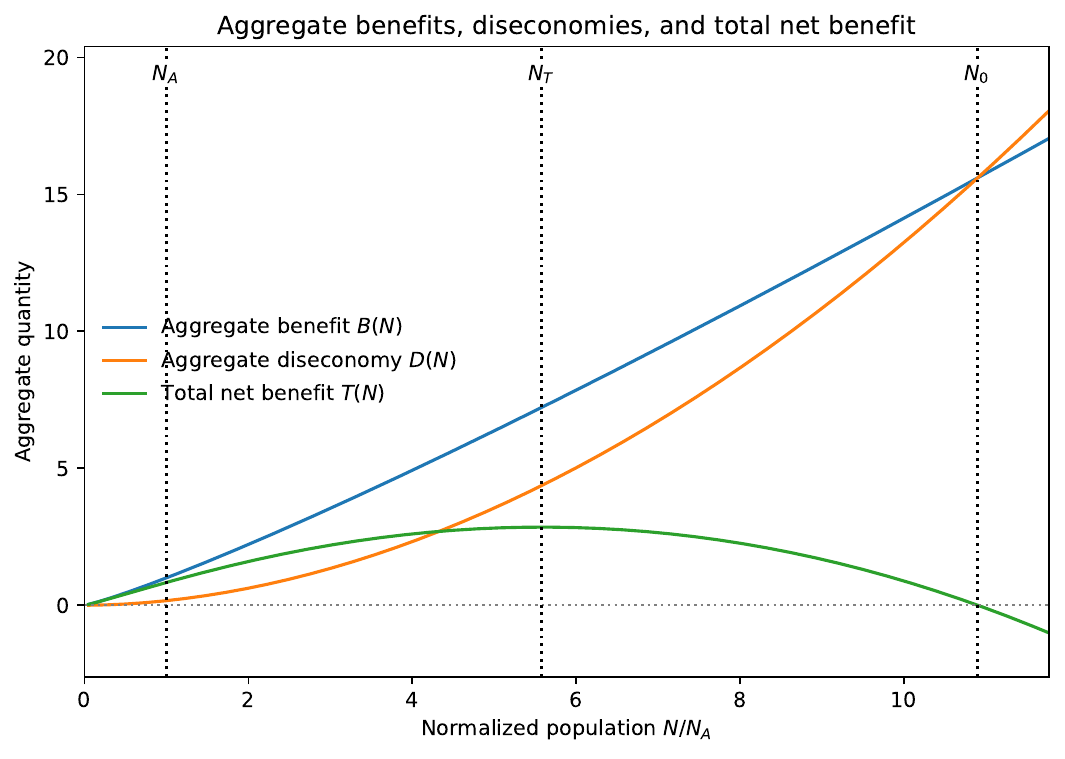}
    \caption{\textbf{Aggregate agglomeration benefits, diseconomies, and total net benefit across the three characteristic urban scales.}
    Aggregate benefits, $B(N)=AN^\beta$, and aggregate diseconomies, $D(N)=CN^\gamma$, both increase superlinearly with population. Total net benefit, $T(N)=B(N)-D(N)$, continues to increase beyond the per-capita optimum $N_A$, reaches its unique maximum at $N_T$, and declines thereafter. At $N_0$, aggregate benefits and diseconomies are equal and total net benefit falls to zero. Population is normalised by $N_A$; the illustration uses $\beta=1.15$ and $\gamma=1.90$.}
    \label{fig:total_three}
\end{figure}

\begin{figure}[htbp]
    \centering
    \includegraphics[width=0.82\linewidth]{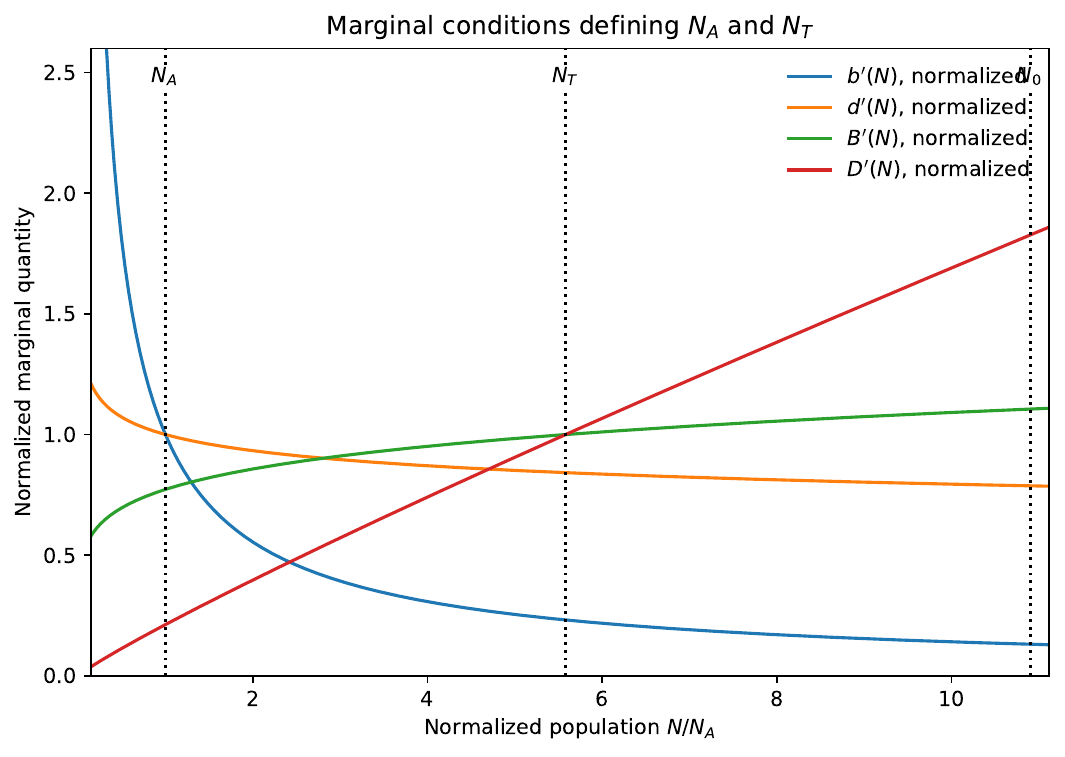}
    \caption{\textbf{Marginal conditions defining the per-capita and total net-benefit maxima.}
    The per-capita optimum $N_A$ occurs where the marginal increase in per-capita benefit, $b'(N)$, equals the marginal increase in per-capita diseconomy, $d'(N)$. The larger total-net-benefit optimum $N_T$ occurs where the marginal increase in aggregate benefit, $B'(N)$, equals the marginal increase in aggregate diseconomy, $D'(N)$. The two equality conditions are distinct, producing separate characteristic population scales. Curves are normalised to make the respective intersections directly comparable. Population is normalised by $N_A$; the illustration uses $\beta=1.15$ and $\gamma=1.90$.}
    \label{fig:marginal_three}
\end{figure}

The implications of the three characteristic urban scales can be seen by expressing population relative to the per-capita net-benefit maximum, $N/N_A$. Figures~\ref{fig:percapita_three}--\ref{fig:marginal_three} use an illustrative parameterisation with $\beta=1.15$ and $\gamma=1.90$, with the coefficients normalised such that $N_A=1$. For these exponents,
\begin{equation}
\frac{N_T}{N_A}\simeq5.58,
\qquad
\frac{N_0}{N_A}\simeq10.90.
\end{equation}
The absolute values are illustrative, but the relative locations of the three scales depend only on $\beta$ and $\gamma$.

Figure~\ref{fig:percapita_three} shows the per-capita quantities. Both per-capita agglomeration benefits,
\begin{equation}
b(N)=AN^{\beta-1},
\end{equation}
and per-capita diseconomies,
\begin{equation}
d(N)=CN^{\gamma-1},
\end{equation}
increase with population. Because $\gamma>\beta$, however, per-capita diseconomies increase more rapidly. Their difference,
\begin{equation}
\psi(N)=b(N)-d(N),
\end{equation}
initially rises with population, reaches its unique maximum at $N_A$, and subsequently declines. Importantly, $\psi(N)$ remains positive well beyond $N_A$ and reaches zero only at the larger break-even population $N_0$. The decline in per-capita net benefit therefore begins substantially before agglomeration benefits are fully offset by diseconomies.

Figure~\ref{fig:total_three} shows the corresponding aggregate quantities. Aggregate benefits and diseconomies both continue to increase superlinearly with population. Total net benefit,
\begin{equation}
T(N)=B(N)-D(N)=AN^\beta-CN^\gamma,
\end{equation}
therefore behaves differently from its per-capita counterpart. Because
\begin{equation}
T(N)=N\psi(N),
\end{equation}
total net benefit can continue to increase even after per-capita net benefit has begun to decline. It reaches its unique maximum at $N_T>N_A$, and only subsequently declines towards zero at $N_0$.

The distinction between $N_A$ and $N_T$ can be seen directly in their marginal conditions (Fig.~\ref{fig:marginal_three}). The per-capita optimum is defined by equality between the marginal changes in per-capita benefits and diseconomies,
\begin{equation}
b'(N_A)=d'(N_A),
\end{equation}
whereas the total-net-benefit optimum is defined by equality between the marginal changes in aggregate benefits and diseconomies,
\begin{equation}
B'(N_T)=D'(N_T).
\end{equation}
Because these are different conditions, the two maxima occur at different population scales. At $N_A$, marginal diseconomies in per-capita terms begin to exceed marginal benefits, so further population growth reduces per-capita net benefit. Aggregate net benefit nevertheless continues to increase until the separate marginal condition defining $N_T$ is reached.

Together, the three figures show that the model separates urban scale into four regimes. For $N<N_A$, both per-capita and total net benefit increase with population. For $N_A<N<N_T$, per-capita net benefit declines while total net benefit continues to increase. For $N_T<N<N_0$, both per-capita and total net benefit decline, although net benefit remains positive. Beyond $N_0$, diseconomies exceed agglomeration benefits and net benefit becomes negative. The point at which further concentration ceases to improve per-capita outcomes can therefore occur substantially before both the population that maximises aggregate net benefit and the population at which net benefit is fully exhausted.

\subsection*{Existing models of finite urban scale as special cases of the general scaling form}

The model developed above generalises a mathematical structure that appears repeatedly in urban economics and urban scaling: an increasing benefit term opposed by a second term that increases more rapidly with urban scale. Writing per-capita net benefit as

\begin{equation}
\psi(N)
=
AN^{\beta-1}
-
CN^{\gamma-1},
\label{eq:general_percapita}
\end{equation}

makes this relationship particularly transparent. If we define

\begin{equation}
p=\beta-1,
\qquad
q=\gamma-1,
\end{equation}

then the model takes the simple form

\begin{equation}
\psi(N)=AN^p-CN^q,
\qquad q>p>0.
\label{eq:general_two_power}
\end{equation}

Several influential models of finite urban scale can be represented as particular restrictions on this general two-power form.

A particularly direct example is Henderson's canonical systems-of-cities formulation \citep{henderson2005urbanization}. In that model, output or wages per worker increase with city employment through localisation economies,

\begin{equation}
W(n)=D h^{\varepsilon} n^\delta,
\end{equation}

while the resource cost of commuting per worker increases as $n^{1/2}$. Total commuting costs are therefore proportional to $n^{3/2}$. Abstracting from the human-capital terms that are constant at a point in time, real income per worker consequently has the form

\begin{equation}
I(n)
=
A n^\delta
-
C n^{1/2}.
\label{eq:henderson}
\end{equation}

Equation~\ref{eq:henderson} is a special case of Eq.~\ref{eq:general_two_power}, with

\begin{equation}
p=\delta,
\qquad
q=\frac{1}{2}.
\end{equation}

Equivalently, expressed in terms of our aggregate scaling exponents,

\begin{equation}
\beta=1+\delta,
\qquad
\gamma=\frac{3}{2}.
\end{equation}

Henderson's condition for a unique finite efficient city size,
\begin{equation}
\delta<\frac{1}{2},
\end{equation}
is therefore exactly equivalent to the condition derived here,

\begin{equation}
\gamma>\beta.
\end{equation}

The connection is substantive rather than merely qualitative. Henderson obtains the adverse exponent from an explicit spatial model in which residents occupy a circular city and commute to a central business district, causing total commuting costs to rise as $n^{3/2}$. Our formulation does not impose this particular spatial mechanism or fix the diseconomy exponent at $3/2$. Instead, $\gamma$ is left unrestricted and can in principle be estimated from any observable aggregate urban diseconomy. Henderson's model is therefore recovered as the particular case $\gamma=3/2$ and $\beta=1+\delta$.

A closely related structure appears in the contemporary urban-growth model of \citet{duranton2023urban}. Their final consumption of incumbent city residents can be written as

\begin{equation}
c(N)
=
A' N^{\sigma+\beta_{\mathrm{DP}}}
-
C' N^{\gamma_{\mathrm{DP}}+\theta},
\label{eq:duranton_puga}
\end{equation}

where the first exponent combines agglomeration mechanisms and the second combines the spatial and congestion components of urban costs. Equation~\ref{eq:duranton_puga} is again of the form

\begin{equation}
AN^p-CN^q,
\end{equation}

with

\begin{equation}
p=\sigma+\beta_{\mathrm{DP}},
\qquad
q=\gamma_{\mathrm{DP}}+\theta.
\end{equation}

A finite maximum consequently requires

\begin{equation}
\gamma_{\mathrm{DP}}+\theta
>
\sigma+\beta_{\mathrm{DP}},
\end{equation}

the same mathematical restriction as $q>p$, or equivalently $\gamma>\beta$, in the present model. Duranton and Puga derive these exponents from a detailed microeconomic model of productivity, human-capital accumulation, housing, commuting and planning regulation. The present approach instead treats the corresponding benefit and diseconomy elasticities as quantities that can be estimated directly from cross-city scaling relationships.

An especially revealing parallel occurs within the urban-scaling literature itself. Building on Bettencourt's model, \citet{sugar2021urban} derive a normalised net-urban-benefit relationship which, for their theoretical value $\alpha=2/3$, can be written as

\begin{equation}
\lambda(x)
=
2x^{1/3}-x^{2/3},
\qquad
x=\frac{G}{G^*},
\label{eq:sugar_kennedy}
\end{equation}

where $\lambda=\mathcal{L}/\mathcal{L}^*$ is normalised net urban benefit and $G/G^*$ measures the balance between socioeconomic output and dissipative infrastructure costs. Algebraically, Eq.~\ref{eq:sugar_kennedy} is again a two-power benefit--cost function,

\begin{equation}
Ax^p-Cx^q,
\end{equation}

with the particular restrictions

\begin{equation}
A=2,
\qquad
C=1,
\qquad
p=\frac{1}{3},
\qquad
q=\frac{2}{3}.
\end{equation}

The important distinction is that $x$ in this formulation is not population itself, but a normalised balance parameter derived from the underlying scaling model. The similarity is nevertheless instructive: the familiar rise-and-fall form of net urban benefit follows from the same general mathematical condition that the adverse term carries the larger exponent.

The present formulation therefore nests these earlier cases within a more general population-scaling framework. Henderson fixes the urban-cost exponent through a specific spatial commuting geometry; Duranton and Puga derive both sides from a microfounded model of productivity and urban costs; and Sugar and Kennedy obtain fixed exponents within a normalised scaling-balance formulation. Here, by contrast, both aggregate components are written directly as independently estimable functions of population,

\begin{equation}
B(N)=AN^\beta,
\qquad
D(N)=CN^\gamma,
\end{equation}

without imposing particular values on either exponent.

This generalisation produces results that are not visible when particular exponents or a single optimisation criterion are imposed. First, the condition for finite urban scale can be stated generally as the empirically testable inequality $\gamma>\beta$. Second, the same two scaling functions generate three distinct population scales,

\begin{equation}
N_A<N_T<N_0,
\end{equation}

corresponding respectively to maximum per-capita net benefit, maximum total net benefit, and equality between aggregate benefits and diseconomies. Third, although the absolute values of these scales depend on the relative magnitude $A/C$, their proportional separation does not:

\begin{equation}
\frac{N_T}{N_A}
=
\left[
\frac{\beta(\gamma-1)}
{\gamma(\beta-1)}
\right]^{\frac{1}{\gamma-\beta}},
\qquad
\frac{N_0}{N_T}
=
\left(
\frac{\gamma}{\beta}
\right)^{\frac{1}{\gamma-\beta}}.
\end{equation}

The contribution is therefore not to replace the classical agglomeration--diseconomy account of finite city size, but to generalise it. Models that produce an inverted-U relationship between urban scale and welfare can be understood as particular cases of a broader class in which benefit and diseconomy processes possess different population elasticities. Once those elasticities are allowed to vary independently, finite urban scale becomes an empirical scaling question, and the model yields not a single ``optimal city size'' but a structured relationship among three characteristic urban scales.

\section{Empirical Results}
\label{sec:empirical}

\subsection{Published urban scaling estimates span the regimes predicted by the model}
\label{subsec:empirical_comparisons}

Published estimates of urban benefits and diseconomies display several instances of the exponent ordering central to the model, $\gamma>\beta$. The comparison is observed for housing costs relative to income, long commuting relative to economic output, selected forms of crime and disease relative to wages, and congestion-related delays and emissions relative to economic indicators. It is not universal: some measured outcomes scale more slowly than the corresponding benefit benchmarks. These contrasts allow the model's characteristic scales to be examined across urban systems while recognizing that different forms of urban diseconomy may exhibit different population-scaling exponents.

For each benefit--diseconomy pair we evaluate the coefficient-free ratios derived in the preceding section:
\begin{equation}
\frac{N_A}{N_0}
=\left(\frac{\beta-1}{\gamma-1}\right)^{1/(\gamma-\beta)},
\label{eq:NA_ratio}
\end{equation}
\begin{equation}
\frac{N_T}{N_0}
=\left(\frac{\beta}{\gamma}\right)^{1/(\gamma-\beta)},
\label{eq:NT_ratio}
\end{equation}
where $N_A$ maximises net benefit per capita, $N_T$ maximises total net benefit, and $N_0$ denotes the break-even population at which aggregate benefits and diseconomies are equal. The ratios are defined here for $\gamma>\beta>1$ and depend only on the estimated exponents; the coefficients of the underlying scaling relationships determine the absolute value of $N_0$.

To provide a common reference scale, Table~\ref{tab:empirical_thresholds} also gives illustrative absolute populations for $N_0=40$ million. This normalisation is anchored to the upper end of observed contemporary city populations, not to an empirical estimate of the break-even scale. The \textit{World Urbanization Prospects 2025} places Jakarta, the world's largest city under the harmonised Degree of Urbanization definition, at approximately 42 million inhabitants in 2025 \citep{un2025wup}. All tabulated absolute values change in direct proportion to a different assumed $N_0$; neither $40$ million nor the resulting $N_A$ and $N_T$ values should be interpreted as estimates of welfare thresholds for the cities represented in these studies. The scaling of $N_A$ and $N_T$ to an $N_0$ of 40 million is a simple illustrative thought experiment, going by the observation that the largest cities on the planet typically are at about 40 million. 

\begin{table*}[t]
\centering
\caption{Characteristic urban scales implied by published
urban scaling exponents. The dimensionless ratios
$N_A/N_0$ and $N_T/N_0$ are determined by the benefit
and diseconomy exponents $\beta$ and $\gamma$,
respectively, and are independent of the scaling
coefficients. $N_A$ denotes the population maximising
net benefit per capita, $N_T$ the population
maximising total net benefit, and $N_0$ the
break-even population at which aggregate benefits
and diseconomies are equal. Illustrative absolute
populations are reported in millions, assuming
$N_0=40$ million. This normalisation approximates
the contemporary upper bound of observed city
population, rather than an empirically estimated
break-even scale. Rows for which $\gamma\leq\beta$
do not satisfy the finite-scale condition and
are indicated by dashes. All characteristic
populations are conditional implications of the
scaling model, not empirical estimates of
optimal city sizes.}
\label{tab:empirical_thresholds}

\small
\setlength{\tabcolsep}{3.5pt}
\renewcommand{\arraystretch}{1.15}

\begin{tabular}{lccccccc}
\hline
&
&
&
&
\multicolumn{2}{c}{Dimensionless ratios}
&
\multicolumn{2}{c}{Illustrative populations} \\
\cline{5-6}
\cline{7-8}
Empirical comparison
& $\beta$
& $\gamma$
& $\gamma-\beta$
& $N_A/N_0$
& $N_T/N_0$
& $N_A$ (million)
& $N_T$ (million) \\
\hline

China: GDP--long commuter count\textsuperscript{1}
& 1.160 & 1.300 & 0.140
& 0.0112 & 0.443
& 0.449 & 17.73 \\

US: bottom-decile income--housing cost\textsuperscript{2}
& 1.010 & 1.110 & 0.100
& $3.9\times10^{-11}$ & 0.389
& $<0.001$ & 15.56 \\

US: top-decile income--housing cost\textsuperscript{2}
& 1.270 & 1.290 & 0.020
& 0.0281 & 0.458
& 1.123 & 18.31 \\

US: wages--serious crime\textsuperscript{3}
& 1.120 & 1.160 & 0.040
& 0.00075 & 0.416
& 0.030 & 16.64 \\

US: wages--AIDS\textsuperscript{3}
& 1.120 & 1.230 & 0.110
& 0.00270 & 0.427
& 0.108 & 17.07 \\

US: wages--aggravated assault\textsuperscript{3}
& 1.120 & 1.130 & 0.010
& 0.00033 & 0.411
& 0.013 & 16.44 \\

US: wages--murder\textsuperscript{3}
& 1.120 & 1.220 & 0.100
& 0.00233 & 0.425
& 0.093 & 17.01 \\

US: wages--motor-vehicle theft\textsuperscript{3}
& 1.120 & 1.320 & 0.200
& 0.00742 & 0.440
& 0.297 & 17.59 \\

US: wages--robbery\textsuperscript{3}
& 1.120 & 1.380 & 0.260
& 0.0119 & 0.448
& 0.475 & 17.92 \\

US: wages--larceny\textsuperscript{3}
& 1.120 & 0.980 & -0.140
& --- & ---
& --- & --- \\

US: wages--congestion delay\textsuperscript{4}
& 1.120 & 1.270 & 0.150
& 0.00449 & 0.433
& 0.180 & 17.30 \\

US: wages--excess  CO$_2$ from congestion\textsuperscript{4}
& 1.120 & 1.262 & 0.142
& 0.00409 & 0.431
& 0.164 & 17.26 \\

US: wages--chlamydia\textsuperscript{5}
& 1.120 & 1.040 & -0.080
& --- & ---
& --- & --- \\

US: wages--gonorrhoea\textsuperscript{5}
& 1.120 & 1.100 & -0.020
& --- & ---
& --- & --- \\

US: wages--syphilis\textsuperscript{5}
& 1.120 & 1.290 & 0.170
& 0.00557 & 0.436
& 0.223 & 17.42 \\

\hline

Australia: household income--housing payments
\textsuperscript{6}
& 1.037 & 1.081 & 0.044
& $1.94\times10^{-8}$ & 0.389
& $<0.001$ & 15.56 \\

Australia: employee income--rent
& 1.045 & 1.068 & 0.024
& $1.58\times10^{-8}$ & 0.388
& $<0.001$ & 15.52 \\

Australia: employee income--commuting
& 1.045 & 1.153 & 0.108
& $1.20\times10^{-5}$ & 0.402
& $<0.001$ & 16.09 \\

Australia: employee income--private-motorised
commuting
& 1.045 & 1.116 & 0.072
& $1.57\times10^{-6}$ & 0.396
& $<0.001$ & 15.85 \\

\hline
\end{tabular}

\vspace{2mm}
\begin{minipage}{0.98\textwidth}
\footnotesize
\textit{Sources and notes.}

\textsuperscript{1}
Huang et al.\ \citep{huang2025longcommuting}.
The long-commuting exponent refers to the number
of commuters travelling at least 60 minutes.

\textsuperscript{2}
Heinrich Mora et al.\ \citep{heinrichmora2021income}.
Income and housing costs are measured separately
for the bottom and top income deciles.

\textsuperscript{3}
Bettencourt et al.\ \citep{bettencourt2007growth}
and Yang et al.\ \citep{yang2019urban}.
The benchmark wage exponent is from Bettencourt
et al.; disaggregated crime exponents are from
Yang et al. Serious crime and AIDS estimates
are from Bettencourt et al.

\textsuperscript{4}
Louf and Barthelemy \citep{louf2014congestion}.
The exponents refer to total congestion-related
delay and excess CO$_2$ emissions attributable
to congestion. The benchmark wage exponent
is from Bettencourt et al.\
\citep{bettencourt2007growth}.

\textsuperscript{5}
Patterson-Lomba et al.\
\citep{pattersonlomba2015sti}.
The reported exponents are from the simpler,
unadjusted scaling model. The benchmark wage
exponent is from Bettencourt et al.\
\citep{bettencourt2007growth}.

\textsuperscript{6}
The Australian household-income--housing-payments
comparison is retained as an ancillary result
from the Census-only analysis. The principal
Australian comparisons use employee income.

Published benefit and diseconomy exponents may
refer to different samples, observation periods,
and estimation specifications. Pairwise comparisons
are illustrative applications of the model and
do not constitute jointly estimated benefit--cost
functions. Australian estimates are from the
present study. The employee-income--rent exponent
difference is not statistically significant
at the 5\% level.

The illustrative absolute populations assume
$N_0=40$ million, motivated by the upper range
of contemporary urban populations
\citep{un2025wup}.
\end{minipage}

\end{table*}

The Chinese comparison combines reported GDP scaling of approximately $1.16$ with a long-commuting exponent of approximately $1.30$ for the full urban sample; the latter is stronger in larger cities \citep{huang2025longcommuting}. For US cities, Heinrich Mora et al.\ report housing-cost exponents that exceed income exponents at both ends of the income distribution: $1.11$ against $1.01$ for the bottom decile and $1.29$ against $1.27$ for the top decile \citep{heinrichmora2021income}. These comparisons are particularly relevant because their income measures distinguish distributional positions rather than treating aggregate output as a sufficient indicator of individual economic gain.

For other US comparisons, we use the benchmark wage exponent $\beta=1.12$ reported by Bettencourt et al.\ \citep{bettencourt2007growth}. Bettencourt et al.\ also report exponents of approximately $1.16$ for serious crime and $1.23$ for new AIDS cases. Yang et al.\ report exponents of $1.13$ for aggravated assault, $1.22$ for murder, $1.32$ for motor-vehicle theft, and $1.38$ for robbery, but approximately $0.98$ for larceny \citep{yang2019urban}. Congestion delay and congestion-related CO$_2$ exhibit exponents of approximately $1.270$ and $1.262$, respectively \citep{louf2014congestion}. For sexually transmitted infections, the corresponding estimates are approximately $1.04$ for chlamydia, $1.10$ for gonorrhoea, and $1.29$ for syphilis \citep{pattersonlomba2015sti}.

These comparisons examine individual pairs of empirically
observed urban benefits and diseconomies, rather than jointly
estimated aggregate benefit and cost functions. In reality,
the total benefits and diseconomies associated with urban
agglomeration arise from multiple processes, each potentially
characterized by different scaling exponents and coefficients.
In particular, commuting distances, crime incidence, and
emissions are not directly expressed in the same welfare units
as wages or GDP. Consequently, the characteristic scales
$N_A$ and $N_T$ reported in Table~\ref{tab:empirical_thresholds}
are conditional implications of applying the two-term model
to individual benefit--diseconomy pairs, rather than estimates
obtained from the combined benefits and costs of an urban
system. They should therefore not be interpreted as empirical
estimates of optimal city sizes. Their significance lies instead
in demonstrating how different observed scaling relationships
translate into the model's characteristic scales, and in
revealing the scaling behavior of the two dimensionless
ratios, $N_A/N_0$ and $N_T/N_0$. Optimal estimates of $N_A$, $N_T$ and $N_0$ will arise from a comprehensive model of total benefits versus total costs, but here we still see in an illustrative way reasonable estimates of realistic sizes emerging from simple pairwise comparisons.

The distinction between $N_A$ and $N_T$ has important
normative implications for urban planning. If urban
systems were explicitly oriented towards maximizing
net benefit per capita, the resulting characteristic
city sizes could be substantially smaller than those
associated with maximizing total net benefit.
Interestingly, our illustrative empirical comparisons
produce values of $N_T$ clustering around 16--18
million inhabitants, well within the range of
contemporary large cities. While this does not establish
that existing urban systems are optimizing total net
benefit, it raises the possibility that prevailing
urban development and planning objectives are more
closely aligned with aggregate rather than per-capita
outcomes (a well-noted and critiqued consequence of utilitarianism that guides much of social, spatial and economic planning). More fundamentally, the framework suggests
that urban size need not be regarded merely as an
outcome of agglomeration processes, but can itself
be considered an explicit planning variable.
Given the prevailing conditions governing agglomeration
benefits and diseconomies, the desirable scale of
an urban system depends critically on what objective
urban policy seeks to maximize.

\subsection{The relative total-net-benefit scale
converges across empirical comparisons}
\label{subsec:empirical_total_net}

The striking feature of
Table~\ref{tab:empirical_thresholds} is the relative
stability of $N_T/N_0$ despite substantial variation
in $\gamma-\beta$. For comparisons satisfying
$\gamma>\beta$, the ratio lies approximately between
$0.39$ and $0.46$. Under the illustrative
normalization $N_0=40$ million, these correspond
to $N_T$ values between approximately 15.5 and
18.3 million inhabitants. Murder, AIDS, robbery,
motor-vehicle theft, congestion delay,
congestion-related CO$_2$ emissions, and syphilis
give values in the narrower range of approximately
16.6--17.9 million. The Chinese long-commuter
comparison yields approximately 17.7 million,
while the US housing comparisons span
approximately 15.6--18.3 million. Although
these illustrative absolute populations depend
on the assumed value of $N_0$, the relative
stability of $N_T/N_0$ across different
empirical scaling relationships is independent
of this normalization.

The clustering follows from the model rather than providing independent evidence of a common absolute urban optimum. From Eq.~\ref{eq:NT_ratio},
\begin{equation}
\ln\!\left(\frac{N_T}{N_0}\right)
=\frac{\ln\beta-\ln\gamma}{\gamma-\beta}.
\label{eq:log_NT_ratio}
\end{equation}
For $\gamma\rightarrow\beta$, the expression has the finite limit
\begin{equation}
\lim_{\gamma\rightarrow\beta}\frac{N_T}{N_0}
=e^{-1/\beta}.
\label{eq:NT_limit}
\end{equation}
For benefit exponents around $1.0$--$1.3$, this limiting ratio varies only modestly. Thus the total-net-benefit maximum is comparatively stable \emph{relative to an assumed break-even population}, even when the two exponents are close or when their difference changes across systems.

The analogous ratio for the per-capita maximum behaves differently:
\begin{equation}
\lim_{\gamma\rightarrow\beta}\frac{N_A}{N_0}
=e^{-1/(\beta-1)}.
\label{eq:NA_limit}
\end{equation}
Because $\beta$ is often close to unity, the denominator $\beta-1$ can be small. The resulting ratio $N_A/N_0$ is highly sensitive to small changes in the exponents and can span many orders of magnitude. This explains why the implied $N_A$ values in Table~\ref{tab:empirical_thresholds} vary far more than the implied $N_T$ values. The extremely small illustrative $N_A$ values for some comparisons should not be read as meaningful predictions for small cities; they expose the fragility of extrapolating the per-capita stationary point from exponents near one.

The published uncertainty intervals are consistent
with this distinction. For US studies, the reported
95\% confidence interval for the wage exponent is
approximately $1.09$--$1.13$
\citep{bettencourt2007growth}, compared with
$1.14$--$1.30$ for murder, $1.18$--$1.29$
for AIDS, $1.27$--$1.38$ for motor-vehicle
theft, and $1.31$--$1.45$ for robbery
\citep{bettencourt2007growth,yang2019urban}.

Because these estimates are not obtained from
a joint statistical model, and their covariance
is unavailable, the published marginal confidence
intervals cannot be propagated as formal joint
confidence intervals for the characteristic scales.
Instead, evaluating the model at the outer
combinations of the reported exponent intervals
provides a sensitivity envelope. Conditional on
$N_0=40$ million, these calculations imply
approximate $N_T$ ranges of
16.31--17.54 million for murder,
16.57--17.48 million for AIDS,
17.11--17.98 million for motor-vehicle theft,
and 17.34--18.35 million for robbery.

As an additional sensitivity comparison,
the covariate-adjusted syphilis exponent reported
by Patterson-Lomba et al.\
\citep{pattersonlomba2015sti} has a 95\%
confidence interval of $[1.212,1.372]$.
Combining this interval with the published
wage interval gives an illustrative $N_T$
range of approximately 16.76--17.94 million.
This calculation uses the adjusted disease
specification, rather than the simpler
unadjusted specification reported in
Table~\ref{tab:empirical_thresholds}.

These calculations should not obscure the importance
of comparisons in which the required exponent
ordering is not supported. Larceny and chlamydia
scale less rapidly than the wage benchmark, as
does the point estimate for gonorrhoea; conversely,
although the aggravated-assault exponent is
slightly greater than the wage exponent, their
reported confidence intervals overlap
\citep{yang2019urban,pattersonlomba2015sti}.
The model's finite-scale regime is conditional
on the relative scaling of benefits and the
diseconomy under consideration. It does not
imply that every urban burden grows more
rapidly than every measure of economic advantage.
The Australian analysis below therefore examines
exponent differences and their uncertainty within
a single geographic system, rather than relying
only on cross-study point estimates.

\subsection{Australian scaling estimates and robustness}
\label{subsec:australia_robustness}

We examined the scaling of employee earnings, rental payments, and commuting distances across 102 Australian Significant Urban Areas (SUAs). The analysis combines 2021 Census population, rent, and distance-to-work observations with 2020--21 administrative employee-income data from the Australian Bureau of Statistics (ABS) \citep{ABS2021ASGS,ABS2021Census,ABS202223Income}. The observation periods are closely aligned, although the Census and financial-year income records do not have identical reference dates. Employee-income totals and numbers of earners were aggregated from Statistical Areas Level 2 (SA2s) to 2021 SUA boundaries using ABS allocation files; mean earnings were computed from these aggregated totals rather than by averaging SA2 means \citep{ABS2021Allocation}. The use of employee income, rather than household income, keeps the benefit proxy closer to labour-market earnings associated with agglomeration.

Aggregate rental payments were reconstructed from the 2021 Census rent distributions (G40), using stated weekly rent-band midpoints, assigning \$950 per week to the open-ended upper band and annualising the estimated weekly total. The principal commuting proxy is the aggregate one-way distance reconstructed from the 2021 Census distance-to-work (DTWP) bands, using distances below 50 km (with nil distance included). Representative distances of 0, 1.25, 6.25, 20, and 40 km were used for the included categories. We also examined the alternative range below 250 km to assess sensitivity to long journeys. Private-motorised commuting distances were estimated from the joint 2021 Census DTWP--method of travel to work (MTWP) cross-tabulation, including car drivers and motorbike/scooter riders but excluding car passengers to avoid double-counting vehicle journeys. These measures represent estimated one-way commuting distances rather than observed journey distances, and do not directly measure transport expenditure or welfare loss.

For each variable, we estimated log--log ordinary least-squares regressions against SUA population, reporting HC3 heteroskedasticity-robust 95\% confidence intervals. Exponent differences were estimated directly on the same cities through regressions of the difference between the logarithms of each burden proxy and employee income on $\ln N_i$, rather than by comparing whether separate confidence intervals overlap. We also evaluated uncertainty using 10,000 paired bootstrap resamples of the 102 SUA-level observations. The aggregate results and paired comparisons are reported in Table~\ref{tab:australia_aggregate}.

\begin{table}[t]
\centering
\caption{Australian aggregate scaling estimates for 102 SUAs. Intervals are HC3-robust 95\% confidence intervals. Paired comparisons report differences between the scaling exponents of the respective burden proxies and total employee income. Commuting is measured in estimated one-way kilometres, not dollars. Private-motorised commuting includes car drivers and motorbike/scooter riders, excluding car passengers.}
\label{tab:australia_aggregate}
\small
\begin{tabular}{lcc}
\hline
Quantity & Exponent & 95\% CI \\
\hline
Total employee income
& 1.0446 & [1.0196, 1.0696] \\
Aggregate annual rent
& 1.0684 & [1.0352, 1.1017] \\
Commuting distance ($<50$ km)
& 1.1528 & [1.1202, 1.1854] \\
Private-motorised distance ($<50$ km)
& 1.1164 & [1.0639, 1.1689] \\
\hline
Paired exponent difference & Estimate & 95\% CI \\
\hline
Rent minus income
& 0.0239 & [-0.0079, 0.0557] \\
Commuting minus income
& 0.1082 & [0.0733, 0.1432] \\
Private-motorised minus income
& 0.0718 & [0.0141, 0.1295] \\
\hline
\end{tabular}
\end{table}

Aggregate employee income scales superlinearly with population, with $\beta_I=1.0446$. Aggregate rent also scales superlinearly, with $\gamma_R=1.0684$, but the paired difference $\gamma_R-\beta_I=0.0239$ is not distinguishable from zero at the 5\% level (HC3 95\% CI $[-0.0079,0.0557]$; $p=0.141$). The paired-bootstrap interval is likewise $[-0.0078,0.0552]$. Although 93.2\% of bootstrap estimates have $\gamma_R>\beta_I$, this frequency is not a posterior probability and the interval includes exponent differences of either sign. Accordingly, the aggregate rent comparison provides suggestive, not conclusive, evidence for the model's exponent ordering.

By contrast, aggregate commuting distance for journeys under 50 km scales with $\gamma_D=1.1528$. Its difference from the employee-income exponent is $0.1082$ (HC3 95\% CI $[0.0733,0.1432]$; $p<0.001$), with a paired-bootstrap interval of $[0.0744,0.1436]$. The private-motorised commuting specification similarly yields an aggregate exponent of $\gamma_{\mathrm{PM}}=1.1164$, with a paired difference from employee income of $0.0718$ (HC3 95\% CI $[0.0141,0.1295]$; $p=0.0147$). For these physical commuting proxies, the evidence of faster aggregate scaling is therefore substantially stronger than for rent.

The uncertainty contrast between $N_A$ and $N_T$ is also evident within the Australian system. The aggregate income--rent point estimates give $N_T/N_0=0.3881$ but $N_A/N_0=1.58\times10^{-8}$. Restricting paired bootstrap draws to the finite-scale regime $\gamma_R>\beta_I>1$ yields 95\% intervals of $[0.3794,0.3964]$ for $N_T/N_0$ and approximately $[3.5\times10^{-15},3.5\times10^{-6}]$ for $N_A/N_0$. Under the purely illustrative normalisation $N_0=40$ million, the first interval corresponds to 15.18--15.86 million around a point value of 15.52 million. Because the exponent ordering is not established for rent across all bootstrap draws, these are \emph{conditional} distributions of model-implied ratios, not confidence intervals for an empirically identified Australian optimal city size. For the under-50-km commuting comparison, $N_T/N_0=0.4022$ with a conditional paired-bootstrap interval of $[0.3941,0.4100]$; the corresponding $N_A/N_0$ remains much less stable.

We examined two substantive sensitivities. First, nine SUAs contain at least one suppressed SA2 employee-income component in 2020--21. For these SUAs, aggregate income was calculated from the available published SA2 components, without treating suppressed values as observed zeros. Consequently, the resulting income totals may be understated. All 102 SUAs were retained in the principal analysis to preserve the full range of city sizes, notably Sydney and Melbourne. A complete-component sensitivity analysis excluding the nine affected SUAs gives a rent--income difference of $0.0459$ (HC3 95\% CI $[-0.0009,0.0926]$; $p=0.055$). This restriction also removes several of the largest metropolitan areas, so the change cannot be attributed to suppression alone, and we do not prefer it to the full-sample estimate.

Second, the commuting result depends on the treatment of longer trips. Extending the distance range from under 50 km to under 250 km changes the aggregate distance exponent from 1.1528 to 1.0611. The corresponding difference from total employee income falls to $0.0165$ (HC3 95\% CI $[-0.0397,0.0727]$; $p=0.565$). Thus, the strong exponent ordering found for shorter commuting journeys is not robust to the inclusion of longer journeys. The 50 km threshold is intended to capture routine
journeys to work within Australia's metropolitan
labour markets, including those of its largest cities,
while distinguishing these from longer-distance commuting. Together, the results indicate why it is necessary to distinguish the particular urban burden being measured, its observational definition, and uncertainty in the estimated exponents. The next subsection examines a further distinction that aggregate comparisons conceal: how much of each exponent reflects the number of participants, and how much reflects the amount borne or received per participant.

\subsection{Extensive and intensive scaling in Australian cities}
\label{subsec:australia_extensive_intensive}

Aggregate economic activity and aggregate urban burdens combine two analytically distinct processes. Their magnitude may increase because the number of earners, renting households, or commuters rises with city size, because the mean amount per participant rises, or because both increase. Consequently, aggregate scaling exponents do not alone reveal how earnings and costs vary for the people or households involved.

Let $E(N)$ be the number of employee-income earners and $\bar{w}(N)$ their mean employee earnings; let $R(N)$ denote renter households with stated weekly rent and $\bar{r}(N)$ their mean annual rent-payment proxy. Aggregate employee income $I(N)$ and rent payments $H_R(N)$ satisfy
\begin{equation}
I(N)=E(N)\bar{w}(N),\qquad
H_R(N)=R(N)\bar{r}(N).
\label{eq:australia_income_rent_identity}
\end{equation}
If the factors follow population-scaling relationships, their aggregate exponents decompose additively:
\begin{equation}
\beta_I=\beta_E+\beta_{\bar{w}},\qquad
\gamma_R=\gamma_{R_n}+\gamma_{\bar{r}}.
\label{eq:australia_decomposition_rent}
\end{equation}
Here $\beta_E$ and $\gamma_{R_n}$ measure the extensive growth of participating units, while $\beta_{\bar{w}}$ and $\gamma_{\bar{r}}$ measure the intensive scaling of earnings and rent per relevant unit. The same identity applies to commuting: writing $D(N)=P_D(N)\bar{d}(N)$ for aggregate distance, the number of persons in the selected DTWP range, and their mean distance gives
\begin{equation}
\gamma_D=\gamma_{P_D}+\gamma_{\bar{d}}.
\label{eq:australia_decomposition_distance}
\end{equation}
An equivalent decomposition applies to private-motorised commuting, where the extensive component counts car drivers and motorbike/scooter riders travelling positive distances below 50 km, and the intensive component measures their mean one-way distance. Because the same 102 SUAs and log-population regressor are used throughout, these decompositions are exact at the regression-slope level, apart from the rounding of displayed estimates.

\begin{table}[t]
\centering
\caption{Extensive and intensive scaling exponents for Australian SUAs. The extensive component refers to the number of employee-income earners, renter households reporting stated weekly rent, or persons in the relevant distance-to-work categories. The intensive component is the corresponding mean employee income, mean rent-payment proxy, or mean one-way commuting distance. Overall commuting includes nil distances; the private-motorised specification includes positive distances below 50 km for car drivers and motorbike/scooter riders. Small discrepancies in sums arise from rounding.}
\label{tab:australia_extensive_intensive}
\small
\begin{tabular}{lccc}
\hline
Quantity & Extensive & Intensive & Aggregate \\
\hline
Employee income
& 1.0153 & 0.0293 & 1.0446 \\
Rental payments
& 1.0025 & 0.0660 & 1.0684 \\
Commuting distance ($<50$ km)
& 1.0345 & 0.1183 & 1.1528 \\
Private-motorised distance ($<50$ km)
& 0.9928 & 0.1236 & 1.1164 \\
\hline
\end{tabular}
\end{table}

Table~\ref{tab:australia_extensive_intensive} shows that the superlinearity of total employee income comprises an almost proportional increase in the number of earners ($\beta_E=1.0153$), together with a smaller positive mean-earnings exponent ($\beta_{\bar{w}}=0.0293$). The rental-payment exponent has a different composition: the number of renter households with stated rent scales approximately proportionally to population ($\gamma_{R_n}=1.0025$), whereas their mean rental-payment proxy increases with exponent $\gamma_{\bar{r}}=0.0660$. Thus, most of the excess of aggregate rental-payment scaling over linearity is associated with the intensive rather than extensive component.

The difference in mean-rent and mean-earnings exponents is
\begin{equation}
\gamma_{\bar{r}}-\beta_{\bar{w}}=0.0366,
\label{eq:australia_mean_rent_income_difference}
\end{equation}
with an HC3 95\% interval of $[0.0113,0.0620]$ ($p=0.0046$) and a paired-bootstrap 95\% interval of $[0.0117,0.0631]$. Unlike the aggregate rent--income exponent difference, this intensive difference is distinguishable from zero. Its larger magnitude can be understood directly by decomposing the aggregate difference:
\begin{equation}
\underbrace{\gamma_R-\beta_I}_{0.02387}
=
\underbrace{(\gamma_{R_n}-\beta_E)}_{-0.01278}
+
\underbrace{(\gamma_{\bar{r}}-\beta_{\bar{w}})}_{0.03665}.
\label{eq:australia_intensive_extensive_contrast}
\end{equation}
The number of renter households with stated rent has a smaller estimated scaling exponent than the number of employee-income earners; this negative extensive difference partly offsets the positive intensive difference. Hence a relatively weak aggregate rent--income comparison can coexist with a clearer difference between the mean amounts measured for the two participating populations.

Commuting shows a stronger intensive gradient under the 50-km restriction. The number of persons represented in the selected distance bands scales with exponent $\gamma_{P_D}=1.0345$, while mean one-way distance scales with exponent $\gamma_{\bar{d}}=0.1183$. Comparing the intensive commuting exponent with mean employee income gives
\begin{equation}
\gamma_{\bar{d}}-\beta_{\bar{w}}=0.0890,
\label{eq:australia_mean_commute_income_difference}
\end{equation}
with an HC3 95\% interval of $[0.0546,0.1234]$ ($p<0.001$) and a paired-bootstrap interval of $[0.0560,0.1248]$. In this case, both differences are positive: the extensive difference is $1.0345-1.0153=0.0192$, and the intensive difference is $0.0890$, adding to the aggregate commuting--income difference of $0.1082$.

The private-motorised commuting comparison exhibits a related pattern. The number of car drivers and motorbike/scooter riders in the selected distance bands scales approximately proportionally to population, with an extensive exponent of $0.9928$, whereas their mean one-way commuting distance scales with an exponent of $0.1236$. The latter exceeds the mean employee-earnings exponent by $0.0943$ (HC3 95\% CI $[0.0562,0.1323]$; $p<0.001$), with a paired-bootstrap interval of $[0.0570,0.1339]$. The negative extensive difference of $-0.0224$ partly offsets this positive intensive difference, giving the aggregate private-motorised commuting--income difference of $0.0718$. Thus, the faster aggregate scaling of private-motorised commuting distances is associated primarily with increasing mean distances rather than a more rapid increase in the number of drivers and riders.

These commuting conclusions are specific to the selected distance range. For distances below 250 km, the overall mean-distance--mean-income exponent difference falls to $0.0028$ (HC3 95\% CI $[-0.0509,0.0565]$; $p=0.918$).

The rent comparison is also sensitive to the observational denominator. In the principal decomposition, the mean rental-payment proxy is calculated from the aggregate rent reconstructed using Census G40 stated-rent categories, divided by the corresponding number of households with stated rents. This ensures consistency between the numerator and denominator. As an alternative, using the 2021 Census Family Records rent bands directly to estimate a frequency-weighted mean gives a mean-rent exponent of $0.0648$. Its paired difference from mean employee earnings is $0.0355$ (HC3 95\% CI $[0.0097,0.0612]$; $p=0.007$). These constructions use different Census record bases, but both support a positive intensive rent--earnings exponent difference. Neither should be mistaken for a precise estimate of rent paid by the same individuals whose earnings are observed.

The extensive--intensive distinction is therefore substantively consequential for interpreting urban scaling. Australian employee income, rent and shorter commuting distances all rise more rapidly than population in aggregate, yet the underlying count and mean components differ substantially. The most clearly supported rent-related result is that the mean rental-payment proxy rises more rapidly with population than mean employee earnings, even though the aggregate rent--income difference remains statistically uncertain. This does not itself demonstrate that particular earners are becoming worse off: the employee-income and rental denominators refer to different populations, and monetary rental payments and physical commuting distances do not together constitute a social welfare function. Rather, it establishes a narrower and important point for the theory: superlinear growth of aggregate urban benefits does not guarantee an equally rapid improvement in intensive economic outcomes relative to the burdens associated with urban size.

\section{Discussion and Conclusions}
\label{sec:conclusion}

The primary contribution of this paper is to connect
the classical urban-economic account of agglomeration
benefits and diseconomies, which motivates models of
finite urban size, with the generalized relationships
described by urban scaling. 

The resulting model identifies three distinct
characteristic population scales when the scaling
exponent of diseconomies exceeds that of benefits,
with $\gamma>\beta>1$: the population $N_A$ at which
net per-capita benefit is maximised, the population
$N_T$ at which total net urban benefit is maximised,
and the population $N_0$ at which aggregate benefits
and diseconomies are equal. These scales are ordered
$N_A<N_T<N_0$. The distinction introduces an
explicitly normative dimension to urban scaling.
Rather than regarding city size solely as an
outcome of agglomeration processes, it becomes
possible to consider urban scale in relation to
the objectives of planning and development.
Maximizing net benefit per capita implies a
smaller characteristic city size than maximizing
total net benefit, even under identical
agglomeration conditions.

A further result is that the relative separation
of these characteristic scales is independent
of the coefficients of benefits and diseconomies.
While their absolute population values depend
on these coefficients, the ratios $N_A/N_0$
and $N_T/N_0$ depend entirely on the scaling
exponents. The model thus separates the
conditions determining the absolute scale of
an urban system from those determining the
relative positions of its characteristic scales.
This distinction is important because it
identifies how different relationships between
agglomeration benefits and diseconomies generate
different implications for finite urban size.

The empirical comparisons illustrate these
relationships using observed scaling exponents
for urban benefits and diseconomies. Although
individual pairings do not constitute complete
urban benefit--cost functions, they generate
characteristic relative scales which, under
the illustrative normalization of $N_0=40$
million, correspond to populations within
the range of contemporary large cities.
The Australian analysis further reveals
important differences between extensive
and intensive scaling. Mean rental payments
increase more rapidly with city population
than mean employee earnings, while a similar
pattern is observed for daily commuting
distances. These
findings demonstrate that superlinear
aggregate scaling cannot be understood
through the number of participating units
alone; the quantities experienced per
participant also exhibit distinct scaling
relationships.

This is where the bridge between urban scaling
and urban economics becomes a more explicitly
normative urban science
\citep{sarkar2024normative,sarkar2026unequal}.
The task is not to infer how large cities
\textit{ought} to be from their scaling exponents
alone, but to determine under what institutional
and distributive conditions further agglomeration
improves people's lives, for whom, and at
whose expense.

\section{Data Availability Statement}

All the data employed for this work is publicly available and has been analyzed using standard Python packages, and the analysis can be reproduced by downloading the data from the public sources mentioned and running the steps of analysis. 

\section{Acknowledgments}
The author thanks Mike Batty and David Levinson for helpful discussions.

\bibliographystyle{SageH}
\bibliography{references.bib}

@article{henderson1974sizes,
  author  = {Henderson, J. Vernon},
  title   = {The Sizes and Types of Cities},
  journal = {American Economic Review},
  year    = {1974},
  volume  = {64},
  number  = {4},
  pages   = {640--656}
}

@article{arnott1979optimal,
  author  = {Arnott, Richard},
  title   = {Optimal City Size in a Spatial Economy},
  journal = {Journal of Urban Economics},
  year    = {1979},
  volume  = {6},
  number  = {1},
  pages   = {65--89},
  doi     = {10.1016/0094-1190(79)90017-2}
}

@article{camagni2013optimal,
  author  = {Camagni, Roberto and Capello, Roberta and Caragliu, Andrea},
  title   = {One or Infinite Optimal City Sizes? In Search of an Equilibrium Size for Cities},
  journal = {The Annals of Regional Science},
  year    = {2013},
  volume  = {51},
  number  = {2},
  pages   = {309--341},
  doi     = {10.1007/s00168-012-0548-7}
}

@article{bettencourt2007growth,
  author  = {Bettencourt, Lu{\'i}s M. A. and Lobo, Jos{\'e} and Helbing, Dirk and K{\"u}hnert, Christian and West, Geoffrey B.},
  title   = {Growth, Innovation, Scaling, and the Pace of Life in Cities},
  journal = {Proceedings of the National Academy of Sciences},
  year    = {2007},
  volume  = {104},
  number  = {17},
  pages   = {7301--7306},
  doi     = {10.1073/pnas.0610172104}
}

@article{bettencourt2013origins,
  author  = {Bettencourt, Lu{\'i}s M. A.},
  title   = {The Origins of Scaling in Cities},
  journal = {Science},
  year    = {2013},
  volume  = {340},
  number  = {6139},
  pages   = {1438--1441},
  doi     = {10.1126/science.1235823}
}

@article{sugar2021urban,
  author  = {Sugar, Lorraine and Kennedy, Christopher},
  title   = {Urban Scaling and the Benefits of Living in Cities},
  journal = {Sustainable Cities and Society},
  year    = {2021},
  volume  = {66},
  pages   = {102617},
  doi     = {10.1016/j.scs.2020.102617}
}

@article{louf2014congestion,
  author  = {Louf, R{\'e}mi and Barthelemy, Marc},
  title   = {How Congestion Shapes Cities: From Mobility Patterns to Scaling},
  journal = {Scientific Reports},
  year    = {2014},
  volume  = {4},
  pages   = {5561},
  doi     = {10.1038/srep05561}
}

@book{sarkar2026unequal,
  author    = {Sarkar, Somwrita},
  title     = {Unequal Agglomerations: Urban Structure, City Size, and Spatial Justice},
  year      = {2026},
  publisher = {Palgrave Macmillan},
  address   = {Singapore},
  doi       = {10.1007/978-981-95-5555-0},
  isbn      = {978-981-95-5555-0}
}

@article{webster2024tiebout,
  author  = {Webster, Chris},
  title   = {Tiebout, Coase and Urban Scaling},
  journal = {The Annals of Regional Science},
  year    = {2024},
  volume  = {73},
  number  = {3},
  pages   = {1125--1147},
  doi     = {10.1007/s00168-024-01272-3}
}

@incollection{gomezlievano2025agglomeration,
  author    = {Gomez-Lievano, Andres and Fragkias, Michail},
  title     = {The Benefits and Costs of Agglomeration: Insights from Economics and Complexity},
  booktitle = {Compendium of Urban Complexity},
  year      = {2025},
  pages     = {43--74},
  publisher = {Springer},
  doi       = {10.1007/978-3-031-82666-5_3}
}

@article{reitemeyer2025urban,
  author  = {Reitemeyer, Fabian and Ribeiro, Fabiano L. and Schwarz, Jan and Gugel, Benjamin and Schacht, Alexander and Rybski, Diego},
  title   = {Minimizing Urban Carbon Emissions and Heat Island Intensity: A Theoretical Study},
  journal = {PLOS ONE},
  year    = {2025},
  volume  = {20},
  number  = {9},
  pages   = {e0330079},
  doi     = {10.1371/journal.pone.0330079}
}

@report{un2025wup,
  author      = {{United Nations}},
  title       = {World Urbanization Prospects 2025: Summary of Results},
  institution = {United Nations, Department of Economic and Social Affairs, Population Division},
  number      = {UN DESA/POP/2025/TR/NO. 12},
  address     = {New York},
  year        = {2025}
}

@article{sarkar2024normative,
  author  = {Sarkar, Somwrita},
  title   = {Normative urban science},
  journal = {Environment and Planning B: Urban Analytics and City Science},
  volume  = {51},
  number  = {5},
  pages   = {1079--1081},
  year    = {2024},
  doi     = {10.1177/23998083241246633}
}

@misc{abs2021census,
  author       = {{Australian Bureau of Statistics}},
  title        = {2021 Census of Population and Housing: General Community Profile DataPacks},
  year         = {2021},
  howpublished = {Australian Bureau of Statistics},
  note         = {Significant Urban Area geography}
}

@article{capello2004beyond,
  author  = {Capello, Roberta and Camagni, Roberto},
  title   = {Beyond Optimal City Size: An Evaluation of Alternative Urban Growth Patterns},
  journal = {Urban Studies},
  volume  = {37},
  number  = {9},
  pages   = {1479--1496},
  year    = {2000},
  doi     = {10.1080/00420980020080221}
}

@incollection{henderson2005urbanization,
  author    = {Henderson, J. Vernon},
  title     = {Urbanization and Growth},
  booktitle = {Handbook of Economic Growth},
  editor    = {Aghion, Philippe and Durlauf, Steven N.},
  volume    = {1B},
  chapter   = {24},
  pages     = {1543--1591},
  publisher = {Elsevier},
  address   = {Amsterdam},
  year      = {2005},
  doi       = {10.1016/S1574-0684(05)01024-5}
}

@article{duranton2023urban,
  author  = {Duranton, Gilles and Puga, Diego},
  title   = {Urban Growth and Its Aggregate Implications},
  journal = {Econometrica},
  volume  = {91},
  number  = {6},
  pages   = {2219--2259},
  year    = {2023},
  doi     = {10.3982/ECTA17936}
}

@article{huang2025longcommuting,
  author  = {Huang, Jie and Fan, Wenying and Wang, Jiaoe and Loo, Becky P. Y. and Qi, Wei},
  title   = {Long commuting, social inequality, and urban development with scaling laws in China},
  journal = {The Innovation Geoscience},
  year    = {2025},
  volume  = {3},
  number  = {3},
  pages   = {100164},
  doi     = {10.59717/j.xinn-geo.2025.100164}
}

@article{heinrichmora2021income,
  author  = {Heinrich Mora, Elisa and Heine, Cate and Jackson, Jacob J. and West, Geoffrey B. and Yang, Vicky Chuqiao and Kempes, Christopher P.},
  title   = {Scaling of urban income inequality in the {USA}},
  journal = {Journal of the Royal Society Interface},
  year    = {2021},
  volume  = {18},
  number  = {181},
  pages   = {20210223},
  doi     = {10.1098/rsif.2021.0223}
}

@article{yang2019urban,
  author  = {Yang, V. Chuqiao and Papachristos, Andrew V. and Abrams, Daniel M.},
  title   = {Modeling the origin of urban-output scaling laws},
  journal = {Physical Review E},
  year    = {2019},
  volume  = {100},
  number  = {3},
  pages   = {032306},
  doi     = {10.1103/PhysRevE.100.032306}
}

@misc{MarquisBarthelemy2026,
  title        = {On the Meaning of Urban Scaling},
  author       = {Marquis, Ulysse and Barth{\'e}lemy, Marc},
  year         = {2026},
  eprint       = {2603.30021},
  archivePrefix = {arXiv},
  primaryClass = {physics.soc-ph},
  url          = {https://arxiv.org/abs/2603.30021}
}

@article{pattersonlomba2015sti,
  author  = {Patterson-Lomba, Oscar and Goldstein, Edward and G{\'o}mez-Li{\'e}vano, Andr{\'e}s and Castillo-Chavez, Carlos and Towers, Sherry},
  title   = {Per-capita incidence of sexually transmitted infections increases systematically with urban population size: a cross-sectional study},
  journal = {Sexually Transmitted Infections},
  year    = {2015},
  volume  = {91},
  number  = {8},
  pages   = {610--614},
  doi     = {10.1136/sextrans-2014-051932}
}

@misc{ABS2021ASGS,
  author       = {{Australian Bureau of Statistics}},
  title        = {{Australian Statistical Geography Standard
                   (ASGS): Edition 3, July 2021--June 2026}},
  year         = {2021},
  publisher    = {Australian Bureau of Statistics},
  url          = {https://www.abs.gov.au/statistics/standards/australian-statistical-geography-standard-asgs/edition-3-july-2021-june-2026}
}

@misc{ABS2021Allocation,
  author       = {{Australian Bureau of Statistics}},
  title        = {{ASGS Edition 3: Allocation Files,
                   Significant Urban Areas 2021}},
  year         = {2021},
  publisher    = {Australian Bureau of Statistics},
  url          = {https://www.abs.gov.au/statistics/standards/australian-statistical-geography-standard-asgs/edition-3-july-2021-june-2026/access-and-downloads/allocation-files}
}

@misc{ABS202223Income,
  author       = {{Australian Bureau of Statistics}},
  title        = {{Personal Income in Australia, 2022--23}},
  year         = {2025},
  publisher    = {Australian Bureau of Statistics},
  note         = {Table 5: Employee income, earners and
                  summary statistics by geography},
  url          = {https://www.abs.gov.au/statistics/labour/earnings-and-working-conditions/personal-income-australia/latest-release}
}

%

\end{document}